\documentclass[12pt,a4paper]{amsart}
\pdfoutput=1
\usepackage{amsmath,latexsym,myharv,eriktheorem,times}
\usepackage{graphicx}%
\def\C{\hbox{\hbox{C\hskip-0.5em\lower-0.1ex
          \hbox{\vrule height1.34ex width0.07em }}\hskip0.50em}}

\def\G{\hbox{\hbox{G\hskip-0.525em\lower-0.081ex
          \hbox{\vrule height1.4ex width0.07em }}\hskip0.50em}}

\def\L{\hbox{I\hskip-0.20em L}}

\def\O{\hbox{\hbox{O\hskip-0.525em\lower-0.095ex
          \hbox{\vrule height1.45ex width0.07em}}\hskip0.50em}}
\def\P{\hbox{I\hskip-0.20em P}}
\def\Q{\hbox{\hbox{Q\hskip-0.525em\lower-0.097ex
          \hbox{\vrule height1.47ex width0.07em}}\hskip0.50em}}
\def\R{\hbox{I\hskip-0.23em R}}
\def\U{\hbox{\hbox{U\hskip-0.45em\lower-0.02ex
     \hbox{\vrule height1.54ex width0.07em}}\hskip0.50em}}

\def\cov{\mbox{cov}}
\def\vol{\mbox{vol}}
\def\volapprox{\overline{\mbox{vol}}}
\def\Filt{{\mathcal F}}
\def\L{{\mathbb L}}
\def\T{{\mathbb T}}
\def\bP{\P}
\def\endpr{\begin{flushright} $\Box$\end{flushright}\vskip .3em \noindent }
\def\proof{\noindent{\sc Proof: }}
\swapnumbers \theoremstyle{dissprop} {\theoremstyle{dissdefi}
\newtheorem{definition}{Definition}[section] }
\newtheorem{assumption}[definition]{Assumption}

\newtheorem{remark}[definition]{Remark}
\begin{document}
\title[Arbitrage Free Term Structure Interpolation]{Arbitrage--Free Interpolation in
Models of Market Observable Interest Rates}
\author{Erik Schl\"ogl}
\address{University of Technology, Sydney, Broadway NSW 2007, Australia}
\email{Erik.Schlogl@uts.edu.au}
\date{
November 22, 2001. School of Finance and Economics, University of
Technology, Sydney, PO Box 123, Broadway, NSW 2007, Australia.
E-mail: Erik.Schlogl@uts.edu.au}
\thanks{The author would like to thank Marek Musiela for helpful
discussions, but claims responsibility for any remaining errors. This is an older, working paper version of the paper published as \mycite{OZ:Schloegl:02a} (see references at the end of the paper).}
\bibliographystyle{kluwplus}
\begin{abstract}
Models which postulate lognormal dynamics for interest rates which
are compounded according to market conventions, such as forward
LIBOR or forward swap rates, can be constructed initially in a
discrete tenor framework. Interpolating interest rates between
maturities in the discrete tenor structure is equivalent to
extending the model to continuous tenor. The present paper sets
forth an alternative way of performing this extension; one which
preserves the Markovian properties of the discrete tenor models
and guarantees the positivity of {\em all} interpolated rates.
\end{abstract}
\maketitle

Practitioners have long priced caps, floors and other interest rate derivative
contracts by Black/ Scholes--like\footnote{cf. \mycite{Bla&Sch:73}.}
formulae, a practice usually attributed to the work of \mycite{Z:Black:76}.
Initially, this approach was very much focused on the pricing of individual contracts, without regard to
the arbitrage relationships between different fixed income instruments. The seminal paper of \mycite{Ho&Lee:86}
instigated further research into models of the entire yield curve and fitting initial data, where the
yield curve was given either terms of zero coupon bond prices\footnote{cf.
such papers as \mycite{Ho&Lee:86} or \mycite{Kar&Lep&Myn&Ros&Vis:91}.} or continuously compounded short or forward
rates. The latter approach was taken by \mycite{Hea&Jar&Mor:92} (HJM), who developed a general framework for
models of the term structure of interest rates, deriving drift conditions for instantaneous forward rates which
must be satisfied in \emph{any} arbitrage--free model. Thus the fundamental objects of these models were mainly
mathematical constructs, not market observables. The first work to move away from this paradigm was
\multicite{San&Son:89}{San&Son:89,San&Son:94a}, who construct a binomial term structure model with the
short rate compounded according to market conventions
(e.g. three--month LIBOR) as the fundamental model variable. Taking the modelling of market
observable interest rates to continuous time and bringing theoretical developments back into line with widespread
market practice, \mycite{TJoF:Milt&Sand&Sond:97}\footnote{See also
\mycite{San&Son:94a} and \mycite{San&Son&Mil:95}.} (MSS) embedded
the pricing of caps and floors by Black/Scholes--like formulae in a consistent HJM framework.
The critical assumption for this result is that relative volatility\footnote{The
relative volatility of a diffusion process $X(t)$ is $\sigma$ if its quadratic variation
is given by $X(t)^2\sigma^2dt$.} of observable market forward
rates such as LIBOR is deterministic.\footnote{For a succinct treatment of the significance of this
assumption, see \mycite{OZ:Rady:97}.}

\mycite{OZ:BGM:97} (BGM) resolved key questions in the
construction of such a model, in particular concerning existence and measure
relationships. Given the deterministic volatility assumption, they explicitly identified forward
LIBOR as lognormal martingales under the forward measure to the
end of the respective accrual periods. Their label \emph{Market Models} seems to have become the
generally accepted descriptor for models which postulate lognormality (under the appropriate probability
measure) for market observable interest rates. The same methodology can be applied to forward swap
rates to derive a model which supports the market practice of
pricing swaptions by a Black/Scholes--type formula, as
demonstrated by \mycite{OZ:Jam:97}. \mycite{OZ:Mil&Nie&San:01} use a similar approach to construct a
model of lognormal futures rates (implied by the prices of interest rate futures), which nests
the lognormal forward LIBOR model.

Starting point of our discussion is the work of \mycite{OZ:MR:97},
who bring important clarity into the construction of lognormal
market models. Besides explicitly deriving the relationships
between forward measures of different maturities, they set up the
model of lognormal forward LIBORs for a discrete set of maturities
(the {\em tenor structure}) first and then extend it to a
continuum of maturities by applying the assumptions of BGM, i.e.
lognormality of all forward rates with compounding period $\delta$
and zero volatility for all zero coupon bonds with time to
maturity less than $\delta$. This makes transparent the fact that
extending the model to continuous tenor in effect stipulates
how the model will interpolate between a discrete set of
maturities. Note that in all cases considered in the present paper the stochastic dynamics of interest rates
are modelled in continuous time. Interpolation occurs in the \emph{maturity dimension} only. Furthermore, the
rates modelled have a fixed maturity (as opposed to a fixed time--to--maturity). In simple terms, the question
addressed is how to determine the continuous time dynamics of, say, the yield for an investment maturing next August
given the continuous time dynamics of three--month forward rates for accrual periods starting in March, June, September
and December of every year.

Specifying lognormal dynamics for all $\delta$--compounded forward
rates, as MSS and BGM do, has the obvious advantage that all
caplets and floorlets on such rates are valued by
Black/Scholes--like formulae. However, this elegance comes at a
price. For one, the no--arbitrage with cash property cannot be
guaranteed, i.e. although the forward LIBORs modelled as lognormal
will always be positive, other rates may not
be.\footnote{\mycite{OZ:MR:97} give an example of this.} Second,
the Markovian properties of the discrete tenor model are lost;
this is a fact which can cause considerable problems in numerical
implementations, including Monte Carlo simulation, as we will
discuss below. Third, one may take the pragmatic point of view
that in reality one observes only a discrete number of rates on
the market, with the rest determined by some interpolation method
of choice. Thus one may prefer a version of the model which
implies an interpolation method for future dates that is
transparent and tractable. As the approach of \mycite{OZ:MR:97} to
the construction of a continuous tenor model makes clear, once
volatilities are specified for continuous tenor, one has the
freedom to arbitrarily interpolate {\em initial} observed market
rates. Thereafter, interpolation is fixed by no arbitrage
conditions and must be evaluated numerically.

The alternative way of extending a lognormal market model from
discrete to continuous tenor proposed here addresses each of these
points. It is also worthy to note that while the practical
relevance of the lognormal market models makes the deterministic
volatility case a natural focus of attention, Musiela and
Rutkowski's forward measure approach to term structure modelling
is valid for much more general volatility specifications, and this
generality carries over to a large part of the results presented
here. In particular, this is true for other models constructed in
discrete tenor, such as the lognormal forward swap rate model of
\mycite{OZ:Jam:97}.

In the present paper, the extension to continuous tenor is initially performed by deterministically
interpolating zero coupon bond prices maturing before the earliest
(future) date in the tenor structure (the ``short bonds''), not only for the initial
term structure, but at any future time as well (section
\ref{short}). At any point in time, interpolation of longer--dated
bonds is determined by no arbitrage conditions (section
\ref{long}). Interpolating short bond prices from the two adjacent discrete tenor LIBORs introduces
volatility for the short bonds and the associated interest rates (section \ref{shortvol}).
The dynamics of all interpolated rates remain
Markovian in the state vector of discrete tenor forward LIBORs
(section \ref{Markov}). The properties of the interpolated rates are discussed in section \ref{cc}.
Section \ref{cap} analyses the implications for the pricing of cap/floor contacts which do not
fit into the discrete tenor structure and section \ref{end} concludes.

\section{Interpolation of Short--dated Bonds}\label{short}
Given a filtered probability space
$(\Omega,\{{\mathcal{F}}_t\}_{t\in[0,T^*]},\bP_{T^*})$ satisfying the usual conditions, let
$\{W_{T^*}(t)\}_{t\in[0,T^*]}$ denote a $d$--dimensional standard
Wiener process and assume that the filtration
$\{{\mathcal{F}}_t\}_{t\in[0,T^*]}$ is the usual
$\bP_{T^*}$--augmentation of the filtration generated by
$\{W_{T^*}(t)\}_{t\in[0,T^*]}$.

The model is set up on the basis of assumptions {\bf (BP.1)} and
{\bf (BP.2)} of \mycite{OZ:MR:97}:
\begin{itemize}
\item[{\bf (BP.1)}] For any date $T\in[0,T^*]$, the price process of a zero coupon
bond $B(t,T)$, $t\in[0,T]$ is a strictly positive special
martingale under $\bP_{T^*}$.
\item[{\bf (BP.2)}] For any fixed $T\in[0,T^*]$, the {\em forward
process}
$$
F_B(t,T,T^*)=\frac{B(t,T)}{B(t,T^*)},\qquad\forall\ t\in[0,T]
$$
follows a martingale under $\bP_{T^*}$.
\end{itemize}
Note that assumption {\bf (BP.2)} means that $\bP_{T^*}$ can be
interpreted as the {\em time $T^*$ forward measure} and implies
that the bond price dynamics are arbitrage--free.

The objects to be modelled on the fixed income
markets are the $\delta$-compounded forward rates defined by
\begin{equation}\label{Ldef}
L(t,T)=\delta^{-1}\left(\frac{B(t,T)}{B(t,T+\delta)}-1\right)
\end{equation}
Since the compounding matches the market convention for rates such
as the London Interbank Offer Rate, $L(t,T)$ is also referred to
as forward LIBOR.

Consider a discrete tenor structure $\T=\{T_0,\ldots,T_N\}$. For notational
simplicity, let $T_0=0$ also be the time origin and
$T_{i+1}-T_i=\delta$ $\forall\ i<N$. The dynamics of the
rates $L(\cdot,T_i)$, $1\leq i<N$, are given as in the discrete
tenor version of the lognormal forward rate model put forth by
\mycite{OZ:MR:97}:\footnote{Note that the focus here is on lognormal forward
LIBOR models for expositional purposes only, because this represents the mainstream of the literature.
The results presented here apply equally well to the discrete tenor versions of other models
of market observable rates, such as the lognormal forward swap rate model of
\mycite{OZ:Jam:97} or extensions such
models to more general volatility functions, such as the model of
level--dependent LIBOR volatilities proposed by \mycite{OZ:AA:98}.}
\begin{equation}\label{domesticL}
dL(t,T_{i})=L(t,T_{i})\lambda(t,T_{i})dW_{T_{i+1}}(t)
\end{equation}
where $W_{T_{i+1}}$ is a $d$-dimensional standard Brownian motion under the time $T_{i+1}$
forward measure $\P_{T_{i+1}}$ and $\lambda:\R\times\T\to\R^d$ is a deterministic function of its
arguments, i.e. each $L(t,T_{i})$ is a lognormal martingale under $\P_{T_{i+1}}$.\footnote{For
details of the model construction, see \mycite{OZ:MR:97}.}

Define the function $$ \eta(t):
[T_0,T_N[\to\{0,\ldots,N\} $$ as $$ \eta(t)=
\max\{i\in\{1,\ldots,N\}|T_{i-1}<t\} $$
Similarly to the extensions to continuous tenor proposed in \mycite{OZ:BGM:97}
and \mycite{OZ:MR:97}, we initially make the following
\begin{assumption}\label{zerovol}
Let volatility be zero
for all zero coupon bonds $B(t_1,t_2)$ maturing by the next
(future) date in the tenor structure $\T$, i.e. $t_1\leq t_2\leq
T_{\eta(t_1)}$.
\end{assumption}
\noindent Consequently, at time $T_i$ the price of the zero
coupon bond maturing in $T_{i+1}$ is given by $$
B(T_i,T_{i+1})=(1+\delta L(T_i,T_i))^{-1} $$ and its dynamics are
deterministic thereafter. Since they are deterministic,
interpolation of interest rates with time to maturity less than
$\delta$ is the same as specifying the evolution of the bond
price, and arbitrary deterministic dynamics can be specified for
$B(t,T_{i+1})$, $T_i<t<T_{i+1}$, with $B(T_{i+1},T_{i+1})=1$. For
obvious reasons, these dynamics should be monotonically increasing
and continuous, which means defining the continuously compounded
short rate as some positive function\footnote{Note that this immediately implies that
all interest rates generated by the model will be positive.}
$r(s,L(T_{\eta(s)-1},T_{\eta(s)-1}))$, such that\footnote{Given Assumption \ref{zerovol}, (\ref{ipspec}) is simply a
restatement of the identity (23) in \mycite{TJoF:Milt&Sand&Sond:97}.}
\begin{equation}\label{ipspec}
\exp\left\{\int_{T_i}^{T_{i+1}}r(s,L(T_{\eta(s)-1},T_{\eta(s)-1}))ds\right\}=1+\delta
L(T_i,T_i)\qquad\forall\ i<N
\end{equation}
Thus $r(t)$ is an $\Filt_{\eta(t)-1}$--measurable random variable.
Consequently,
$$
B(t,T_{\eta(t)})=\exp\left\{-\int_t^{T_{\eta(t)}}r(s,L(T_{\eta(s)-1},T_{\eta(s)-1}))ds\right\}
$$ Note that while the dynamics of the abstract variable $r$ are
in general discontinuous at each $T_i$, any rate associated with a
compounding period greater than length zero will have continuous
dynamics.
\begin{remark}
In a term structure model with zero volatility for
zero coupon bonds $B(t,T_{\eta(t)})$ for all $T_0\leq t\leq T_N$, the
risk neutral and rolling spot LIBOR measures are identical and the
short rate interpolation (\ref{ipspec}) is arbitrage free.
\end{remark}
\proof The risk neutral equivalent martingale measure is
conventionally defined as the measure associated with taking the
(continuously compounded) savings account as the numeraire asset,
i.e. under this measure all assets discounted by the savings
account are martingales. The savings account $\beta(t)$ defined by
the interpolation is
\begin{eqnarray}
\beta(t) & = &
\exp\left\{\int_{0}^{t}r(s,L(T_{\eta(s)-1},T_{\eta(s)-1}))ds\right\}\nonumber\\
& = & \left(\prod_{i=0}^{\eta(t)-2}(1+\delta L(T_i,T_i))\right)
\exp\left\{\int_{T_{\eta(t)-1}}^{t}r(s,L(T_{\eta(s)-1},T_{\eta(s)-1}))ds\right\}\label{savacc}
\end{eqnarray}
For the rolling spot LIBOR measure as introduced by
\mycite{OZ:Jam:97}, the numeraire is a roll--over strategy in
which money is invested and subsequently reinvested at spot LIBOR.
Investing one monetary unit in $T_i$ at the spot LIBOR
$L(T_i,T_i)$ is the same as buying $k$ zero coupon bonds
$B(T_i,T_{i+1})$, where $k=B(T_i,T_{i+1})$. Thus in continuous
tenor the value at time $t$ of the roll--over strategy is
\begin{equation}\label{spotnum}
\left(\prod_{i=0}^{\eta(t)-2}(1+\delta
L(T_i,T_i))\right)\frac{B(t,T_{i+1})}{B(T_i,T_{i+1})}
\end{equation}
Since $r(s,L(T_{\eta(s)-1},T_{\eta(s)-1}))$ is
$\Filt_{T_i}$--measurable for $T_i\leq s<T_{i+1}$, we have by
construction that $$
\frac{B(t,T_{i+1})}{B(T_i,T_{i+1})}=\exp\left\{\int_{T_{\eta(t)-1}}^{t}r(s,L(T_{\eta(s)-1},T_{\eta(s)-1}))ds\right\}
$$ for $T_i\leq t\leq T_{i+1}$. Thus the numeraire processes are
identical, which means that the associated martingale measures
must be\footnote{In a continuous tenor model, the market is
(dynamically) complete, therefore the martingale measure for a
given numeraire is unique. For an alternative proof, cf. Lemma 3.3 in
\mycite{OZ:Schloegl:02}.}.

The interpolation defined by (\ref{ipspec}) is consistent with no
arbitrage if the savings account given by (\ref{savacc})
discounted by a numeraire asset is a martingale under the
martingale measure associated with this numeraire. It is
sufficient to verify this for one measure and by the above
argument, this condition is trivially fulfilled since the value of
the savings account is identically equal to the value of the spot
LIBOR roll--over strategy. \endpr

In other words, since at any time $t$ the dynamics of the shortest
remaining bond to a discrete tenor date $B(t,T_{\eta(t)})$ are
consistent with the deterministic interpolation for the
continuously compounded short rate, all martingale properties (and
thus no arbitrage) are preserved.

The observation that the roll--over strategy in spot LIBOR
corresponds to a roll-over strategy in the shortest remaining bond
to a discrete tenor date also leads us to the following
\begin{remark}\label{paste}
{\em Conditional} on the information at time $T_i$, for all
$\Filt_{T_{i+1}}$--measurable events, the rolling spot LIBOR
measure $\P_0$ is identical to the forward measure of maturity
$T_{i+1}$, i.e. $$
\P_0\{A|\Filt_{T_i}\}=\P_{T_{i+1}}\{A|\Filt_{T_i}\}\qquad\forall\
A\in\Filt_{T_{i+1}} $$ Thus $\P_0$ can be interpreted as ``pasting
together'' a sequence of conditional forward measures. This is
valid for any arbitrage free term structure model.
\end{remark}
\proof This is an immediate consequence of the fact that the time
$t$ value of the spot LIBOR roll--over strategy (\ref{spotnum})
can be written, independently of how the discrete tenor model is
extended to continuous tenor, as a $T_{\eta(t)-1}$--measurable
factor times $B(t,T_{\eta(t)})$, the numeraire of the time
$T_{\eta(t)}$ forward measure.\footnote{For a more formal statement
of this proof, see \mycite{OZ:Schloegl:02}.} \endpr

\section{Interpolation of Longer--dated Bonds}\label{long}
\begin{remark}\label{myspec}
Given a discrete tenor model and the interpolation of the
short--dated zero coupon bonds $B(t,T_{\eta(t)})$, the continuous
tenor model is completely specified.
\end{remark}
\proof Consider a completely arbitrary pair of time points $t_1,
t_2$, $0\leq t_1<t_2\leq T_N$. The time $t_1$ price of a zero
coupon bond maturing in $t_2$ is given by
\begin{equation}
B(t_1,t_2)=B(t_1,T_{\eta(t_1)})\left(\prod_{i=\eta(t_1)}^{\eta(t_2)-1}(1+\delta
L(t_1,T_i))^{-1}\right) \frac{B(t_1,t_2)}{B(t_1,T_{\eta(t_2)})}
\end{equation}
where the short--dated bond $B(t_1,T_{\eta(t_1)})$ is given by the
interpolation and the bond price quotient
on the far right by the no--arbitrage requirement that $$
\frac{B(t_1,t_2)}{B(t_1,T_{\eta(t_2)})}=E_{T_{\eta(t_2)}}[B(t_2,T_{\eta(t_2)})^{-1}|\Filt_{t_1}]
$$ Given a choice of interpolation (\ref{ipspec}), we can
determine a function $g$ such that
\begin{equation}\label{longbond}
\frac{B(t_1,t_2)}{B(t_1,T_{\eta(t_2)})}=E_{T_{\eta(t_2)}}[g(L(T_{\eta(t_2)-1},T_{\eta(t_2)-1}))|\Filt_{t_1}]
\end{equation}
In the lognormal case, the right hand side can easily be evaluated
numerically and depends only on $L(t_1,T_{\eta(t_2)-1})$ and
deterministic volatilities. \endpr

When intermediate rates with LIBOR--type compounding are linearly interpolated by day--count fractions,
(\ref{longbond}) becomes very tractable for any model satisfying the assumptions {\bf (BP.1)} and {\bf (BP.2)}. Define
$r(s,L(T_{\eta(s)-1},T_{\eta(s)-1}))$ implicitly by setting
\begin{equation}\label{meth1}
\exp\left\{\int_{t}^{T_{i+1}}r(s,L(T_{\eta(s)-1},T_{\eta(s)-1}))ds\right\}=1+(T_{i+1}-t)
L(T_i,T_i) \qquad\forall\ T_i\leq t<T_{i+1}
\end{equation}
i.e. $$
r(s,L(T_i,T_i))=\frac{L(T_i,T_i)}{1+(T_{i+1}-s)L(T_i,T_i)} $$ Then
$$
B(t_1,T_{\eta(t_1)})=(1+(T_{\eta(t_1)}-t_1)L(T_{\eta(t_1)-1},T_{\eta(t_1)-1}))^{-1}
$$ and
\begin{eqnarray*}
\frac{B(t_1,t_2)}{B(t_1,T_{\eta(t_2)})} & = &
E_{T_{\eta(t_2)}}[B(t_2,T_{\eta(t_2)})^{-1}|\Filt_{t_1}]\\ & = &
1+(T_{\eta(t_2)}-t_2)E_{T_{\eta(t_2)}}[L(T_{\eta(t_2)-1},T_{\eta(t_2)-1})|\Filt_{t_1}]\label{bmeth1}\\
& = & 1+(T_{\eta(t_2)}-t_2)L(t_1,T_{\eta(t_2)-1})
\end{eqnarray*}
Note that by remark \ref{myspec}, if interpolation of short--dated
bonds is specified for all times, there is no modelling freedom to
interpolate longer--dated bonds, including initial model inputs.
In addition to being intuitively appealing, linear interpolation
by day--count fractions has the added attraction of providing one
consistent interpolation method for all times and all maturities.
Other popular interpolation methods, such as loglinear
interpolation of discount factors or linear interpolation of
continuously compounded yields, do not have this property.

\section{Introducing Volatility for the Short Bonds}\label{shortvol}
In some applications, setting short bond volatilities to zero may be unsatisfactory.
Assumption \ref{zerovol} can certainly be relaxed in any way which satisfies the arbitrage
constraints. Remark \ref{myspec} remains valid, i.e. to extend the model to continuous tenor it is
sufficient to specify the dynamics of the short bonds. However, when the short bond dynamics are stochastic,
the constraint given by equation (\ref{ipspec}) is replaced by the more general formulation given in terms of
instantaneous forward rates in equation (23) of \mycite{TJoF:Milt&Sand&Sond:97}:
\begin{equation}\label{meth2}
\exp\left\{\int_{T_i}^{T_{i+1}}f(T_i,s)ds\right\}=1+\delta
L(T_i,T_i)\qquad\forall\ i<N
\end{equation}
where $f(T_i,s)$ is the instantaneous forward rate at time $T_i$ for maturity $s$.

One tractable and intuitively appealing way to introduce volatility for the interpolated short bonds is to
make the interpolated rates dependent on the closest remaining forward LIBOR
$L(t_1,T_{\eta(t_1)})$. Set
$$
B(t_1,T_{\eta(t_1)})^{-1}
=1+(T_{\eta(t_1)}-t_1)(\alpha(t_1)L(T_{\eta(t_1)-1},T_{\eta(t_1)-1})
+(1-\alpha(t_1))L(t_1,T_{\eta(t_1)}))
$$
where
$$
\begin{array}{rr}
\displaystyle\lim_{\Delta\searrow0}\alpha(T_i+\Delta)=1 \\[0.7em]
\displaystyle\lim_{\Delta\nearrow T_{i+1}-T_i}\alpha(T_i+\Delta)=0
\end{array}\qquad\quad\forall\quad i=0,\ldots,N-1
$$
e.g.
$$
\alpha(t)=\frac{T_{\eta(t)}-t}{T_{\eta(t)}-T_{\eta(t)-1}}
$$
As in section \ref{long}, the prices of bonds with longer maturities must satisfy
\begin{eqnarray*}
\frac{B(t_1,t_2)}{B(t_1,T_{\eta(t_2)})}&=&
E_{T_{\eta(t_2)}}\left[\left.\frac{B(t_2,t_2)}{B(t_2,T_{\eta(t_2)})}\right|\Filt_{t_1}\right]\\
&=&1+(T_{\eta(t_2)}-t_2)\left(\alpha(t_2)E_{T_{\eta(t_2)}}
\left[L(T_{\eta(t_2)-1},T_{\eta(t_2)-1})|\Filt_{t_1}\right]\right.\\
&&\qquad\left.+(1-\alpha(t_2))E_{T_{\eta(t_2)}}\left[L(t_2,T_{\eta(t_2)})|\Filt_{t_1}\right]\right)
\end{eqnarray*}
Using the expected value of $L(t,T_j)$ under $\bP_{T_j}$ given in \mycite{OZ:Rutkowski:97}:
\begin{eqnarray}
&=& 1+(T_{\eta(t_2)}-t_2)\Biggl(\alpha(t_2)L(t_1,T_{\eta(t_2)-1})+(1-\alpha(t_2))L(t_1,T_{\eta(t_2)})\nonumber\\
&&\qquad\cdot\underbrace{\left(1+
\frac{(T_{\eta(t_2)+1}-T_{\eta(t_2)})L(t_1,T_{\eta(t_2)})\left(\exp\left\{\int_{t_1}^{t_2}
\lambda^2(s,T_{\eta(t_2)})ds\right\}-1\right)}{1+(T_{\eta(t_2)+1}-T_{\eta(t_2)})L(t_1,T_{\eta(t_2)})}
\right)}_{\mbox{\scriptsize ``correction factor''}}\Biggr)\label{corrfac}
\end{eqnarray}
The ``correction factor'' will usually be quite close to one, unless volatility $\lambda$ is very high or
time to maturity $t_2-t_1$ is very long. Thus interpolation remains essentially linear.

Note that since the short bonds are no longer deterministic,
we also need to evaluate the appropriate expectation to derive
$B(t,T)$ for $t<T<T_{\eta(t)}$:
\begin{eqnarray*}
\frac{B(t,T)}{B(t,T_{\eta(t)})}&=&
E_{T_{\eta(t)}}\left[\left.\frac{B(T,T)}{B(T,T_{\eta(t)})}\right|\Filt_{t}\right]\\
&=& 1+(T_{\eta(t)}-T)\Biggl(\alpha(T)L(t,T_{\eta(t)-1})+(1-\alpha(T))L(t,T_{\eta(t)})\\
&&\cdot\left(1+
\frac{(T_{\eta(t)+1}-T_{\eta(t)})L(t,T_{\eta(t)})\left(\exp\left\{\int_{t}^{T}
\lambda^2(s,T_{\eta(t)})ds\right\}-1\right)}{1+(T_{\eta(t)+1}-T_{\eta(t)})L(t,T_{\eta(t)})}
\right)\Biggr)
\end{eqnarray*}

\section{Forward Measures and the Markov Property}\label{Markov}
Both \mycite{TJoF:Milt&Sand&Sond:97} and \mycite{OZ:BGM:97} set up the model of
lognormal forward LIBORs in continuous tenor, i.e. equation (\ref{domesticL}) applies
to all maturities $T\in(0,T^*-\delta]$:
$$
dL(t,T)=L(t,T)\lambda(t,T)dW_{T+\delta}(t)
$$
with $\lambda:\R\times(0,T^*]\to\R^d$ a deterministic function of its
arguments, i.e. each $L(t,T)$ is a lognormal martingale under $\P_{T+\delta}$ for
a continuum of maturities up to the time horizon $T^*$. This has the advantage that all
caps and floors on $\delta$--compounded rates will be priced by the Black/Scholes--type
formulae favoured by market practitioners.

The main disadvantage of this approach is revealed when one attempts to price instruments
for which closed form solutions are unavailable. This is already the case if the cashflow
underlying a derivative does not fit neatly into a tenor structure of $\delta$--compounded
rates and even more so for many popular interest rate exotics. The problem is that the
continuous tenor model is infinite--dimensional even if the driving Brownian motion is
one--dimensional. That is, the Markovian state variable for such a model is the entire yield
curve.\footnote{The term ``yield curve'' generally denotes the term structure interest rates
for a continuum of maturities up to the time horizon. It can be represented in several
equivalent ways, for example by the curve of all continuously compounded yields at time $t$,
$y_t:(t,T^*]\to\R$ with $y_t(T)$ defined by
$$
B(t,T)=\exp\{-y_t(T)\cdot T\}
$$
} This causes considerable difficulties for all types of numerical methods. Even the method
of last resort for high--dimensional problems, Monte Carlo simulation, cannot handle
infinite--dimensional state variables.

From the start, it has been part of the ``folk wisdom'' on these models that they do not
permit a finite dimensional representation. \mycite{OZ:BGM:97} derive the dynamics of the
continuously compounded short rate under a given set of assumptions on the volatility
structure and show that they are highly path--dependent and thus not Markovian in any finite set
of state variables. Recently, \mycite{OZ:Corr:00} formalised this conjecture and showed
that when the model is driven by a one--dimensional Brownian motion, a finite--dimensional
representation can only exist in the trivial case of zero volatility. He also derives
similarly restrictive conditions on the finite--dimensional representability
of continuous tenor LIBOR models driven
by a multidimensional Brownian motion. These conditions, though stopping short of precluding
finite--dimensional cases, indicate that such cases are not likely to be practically
useful if they do exist.

The extensions of the discrete tenor model to continuous tenor proposed in the
previous sections provide a way around this problem. Since the interpolated rates are
specified as functions of the discrete tenor rates, the Markovian structure of the
discrete tenor model is preserved. This structure can be characterised as follows:

\begin{remark}\label{remarkov}
In the discrete tenor lognormal forward LIBOR model, consider the
dynamics of a rate $L(\cdot,T_i)$ under some forward measure\footnotemark
$\P_{T_j}$. These dynamics are Markovian in a state variable
vector consisting of $n=|j-1-i|+1$ rates $L(\cdot,T_k)$ with
$\min(i,j-1)\leq k\leq\max(i,j-1)$.
\end{remark}
\footnotetext{Note that the Markov property depends on the probability measure
under consideration.}
\proof \mycite{OZ:MR:97} show that the relationship between
forward measures $\P_{T_k}$ and $\P_{T_{k+1}}$ is given by the
Radon/Nikodym derivative $$
\left.\frac{d\P_{T_k}}{d\P_{T_{k+1}}}\right|_{\Filt_t}
=\exp\left\{\int_0^t\gamma(u,T_k,T_{k+1})dW_{T_{k+1}}(u)
-\frac12\int_0^t\gamma^2(u,T_k,T_{k+1})du\right\} $$ where
$W_{T_{k+1}}$ is a Brownian motion under $\P_{T_{k+1}}$ and $$
\gamma(t,T_k,T_{k+1})=\frac{\delta L(t,T_k)}{1+\delta
L(t,T_k)}\lambda(t,T_k) $$ with $\lambda(t,T_k)$ the
(deterministic) volatility of $L(t,T_k)$. Thus $$
dW_{T_k}(t)=dW_{T_{k+1}}(t)-\frac{\delta L(t,T_k)}{1+\delta
L(t,T_k)}\lambda(t,T_k)dt $$ Since the dynamics of $L(\cdot,T_k)$
are given by $$ dL(t,T_k)=L(t,T_k)\lambda(t,T_k)dW_{T_{k+1}}(t) $$
the joint dynamics of all rates $L(\cdot,T_k)$ with
$\min(i,j-1)\leq k\leq\max(i,j-1)$ under $\P_{T_{\max(i+1,j)}}$
are
\begin{equation}\label{back}
d\L(t)=\Lambda(t,\L(t))dW_{T_{\max(i+1,j)}}(t)-\Psi(\L(t))\ell(\L(t))dt
\end{equation}
where $$ \L(t) = \left(\begin{array}{c} L(t,T_{\min(i,j-1)})\\
L(t,T_{\min(i,j-1)+1})\\ \vdots\\ L(t,T_{\max(i,j-1)})\end{array}
\right) $$ If $d$ is the dimension of the driving Brownian motion,
$\Lambda(t,\L(t))$ is an $n\times d$ matrix with $$
\Lambda_{hk}(t,\L(t)) =
L(t,T_{\min(i,j-1)-1+h})\lambda_k(t,T_{\min(i,j-1)-1+h}) $$
$\Psi(t,\L(t))$ is an $n\times n$ matrix with $$
\Psi_{hk}(t,\L(t)) = \left\{\begin{array}{rl}
L(t,T_{\min(i,j-1)-1+h})\lambda(t,T_{\min(i,j-1)-1+h})\lambda(t,T_{\min(i,j-1)-1+k})
& \mbox{ if } k>h\\ 0 & \mbox{ otherwise}\end{array}\right. $$
Finally, $\ell(\L(t))$ is an $n$--dimensional vector with $$
\ell_k(\L(t)) = \frac{\delta L(t,T_{\min(i,j-1)-1+k})}{1+\delta
L(t,T_{\min(i,j-1)-1+k})} $$ Conversely, under
$\P_{T_{\min(i+1,j)}}$ we have
\begin{equation}\label{fwd}
d\L(t)=\Lambda(t,\L(t))dW_{T_{\min(i+1,j)}}(t)+\Psi'(\L(t))\ell(\L(t))dt
\end{equation}
with
$$
\Psi'_{hk}(t,\L(t)) = \left\{\begin{array}{rl}
L(t,T_{\min(i,j-1)-1+h})\lambda(t,T_{\min(i,j-1)-1+h})\lambda(t,T_{\min(i,j-1)-1+k})
& \mbox{ if } 1<k\leq h\\ 0 & \mbox{ otherwise}\end{array}\right. $$
By the Markov property of Ito
diffusions, the dynamics given by (\ref{back}) resp.\ (\ref{fwd})
are Markov in $\L(t)$.\endpr
This result does not rely on the assumption that
the LIBOR volatilities $\lambda(t,T_k)$ are deterministic functions of their arguments.
Rather, if the $\lambda(t,T_k)$ are level dependent on some or all
rates in $\L(t)$, the Markov property still holds. Also, by
remark \ref{paste}, the above result implies that under the rolling spot LIBOR measure, any rate
$L(\cdot,T_i)$ is Markov in the state variable vector
$\{L(t,T_k)|0\leq k\leq i\}$.
\begin{figure}[tb]
\begin{minipage}[t]{0.445\textwidth}
\includegraphics[width=.99\textwidth]{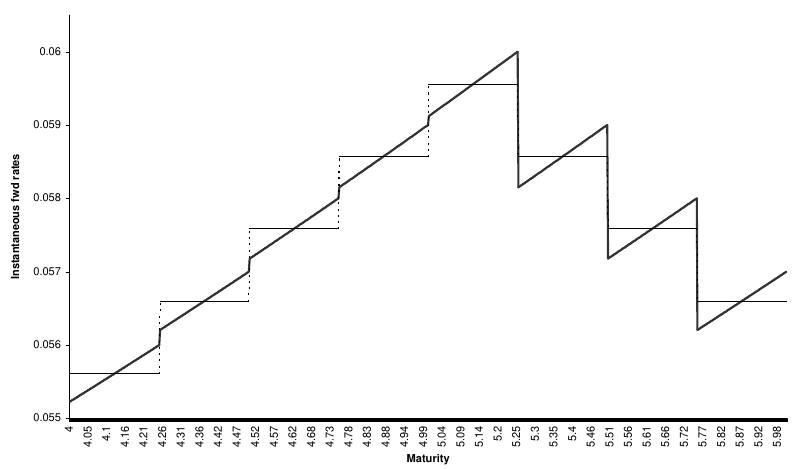}
\caption{}\label{daycountTS}
\end{minipage}\hfill\begin{minipage}[t]{0.445\textwidth}
\includegraphics[width=.99\textwidth]{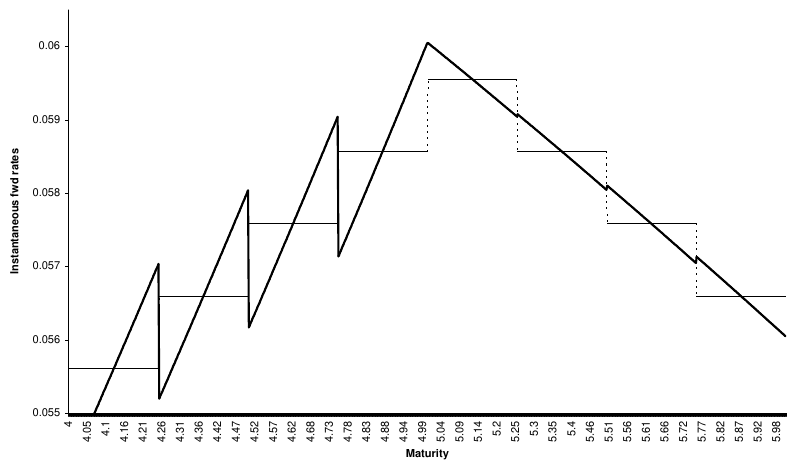}
\caption{}\label{shortvolTS}
\end{minipage}\\[1.2ex]
\centerline{Interpolation by daycount fractions vs. interpolation with short bond volatility}
\end{figure}

\section{A Look at the Interpolated Interest Rates}\label{cc}
\begin{figure}[tb]
\centerline{\includegraphics[width=.95\textwidth]{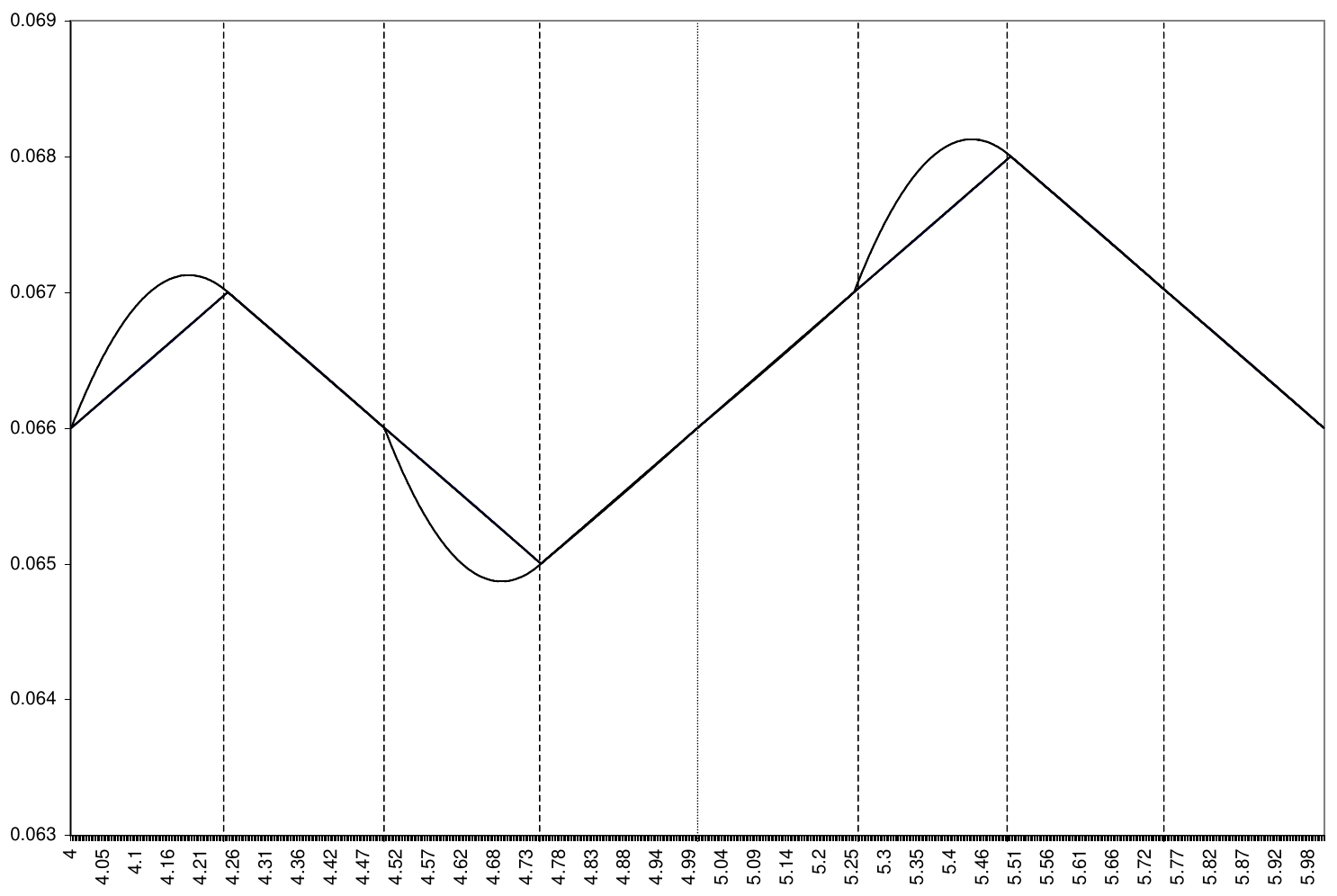}}
\caption{Interpolated forward LIBORs}\label{LIBORTS}
\end{figure}
This section illustrates some of the properties of the
interpolated interest rates and contrasts the two interpolation
methods proposed in the previous sections. The method described
in section \ref{long} will be referred to as method 1 or
\emph{interpolation by daycount fractions}; the one introduced in
section \ref{shortvol} will be labelled method 2 or
\emph{interpolation with short bond volatility}. For the inputs
used to produce each of the plots, please refer to table
\ref{params} in the appendix. The algorithm from
\mycite{FMMA:BMS:98} was used to generate sample paths for the
evolution of the term structure.
\subsection{Term Structures}
Figures \ref{daycountTS} and \ref{shortvolTS} show how the two
methods interpolate instantaneous forward rates. For reference,
the stepwise constant rates resulting from loglinear
interpolation of zero coupon bond prices are included in each
plot. The instantaneous forward rates jump when one moves from
one $\delta$--compounding period to the next. From equations
(\ref{meth1}) and (\ref{meth2}), this is to be expected, as the
maturities of the LIBORs underlying the interpolation are
different on each interval. Instantaneous forward rates on
$f(0,T)$ are determined on each interval $T\in(T_i,T_{i+1})$ by
calculating
$$
-\partial_T\ln\frac{B(0,T)}{B(0,T_{\eta(T)})}
$$
using (\ref{bmeth1}) for method 1 and (\ref{corrfac}) for method
2.

Instantaneous forward rates give the most disaggregate
representation of the term structure and therefore effectively
demonstrate the implications of the choice of interpolation
method. However, one should not judge the usefulness of a
particular method from this perspective alone. At first glance
the ``sawtooth'' patterns of the interpolated rates for falling
term structures in method 1 and rising term structures in method
2 seem to represent serious drawbacks to using either method, but
one should keep in mind that the instantaneous forward rates are
also purely mathematical artifacts, while the objects being
modelled are market observable rates such as LIBOR. Forward
LIBORs are interpolated by
\begin{eqnarray}
L(t,T) & = & \frac1{\delta}\left(\frac{B(t,T)}{B(t,T+\delta)}-1\right)\nonumber\\
& = & \frac1{\delta}\left(\frac{B(t,T)}{B(t,T_{\eta(T)})}
\frac{B(t,T_{\eta(T)})}{B(t,T_{\eta(T)+1})}\frac{B(t,T_{\eta(T)+1})}{B(t,T+\delta)}-1\right)
\nonumber\\
& = & \frac1{\delta}\left(\frac{B(t,T)}{B(t,T_{\eta(T)})}
(1+\delta
L(t,T_{\eta(T)}))\frac{B(t,T_{\eta(T)+1})}{B(t,T+\delta)}-1\right)\label{ipLIBOR}
\end{eqnarray}
For method 1 this becomes
\begin{eqnarray*}
L(t,T) & = & \frac1{\delta}\bigl((1+(T_{\eta(T)}-T)L(t,T_{\eta(T)-1}))(1+\delta L(t,T_{\eta(T)}))\\
&&
\qquad\cdot(1+(T_{\eta(T)+1}-(T+\delta))L(t,T_{\eta(T)}))^{-1}-1\bigr)
\end{eqnarray*}
Figure \ref{LIBORTS} shows interpolated forward three--month
LIBORs for a continuum of start dates between four and six years
forward. The piecewise linear graph is actually two plots so
close as to be indistinguishable: Interpolation by daycount
fractions and loglinear interpolation of zero coupon bond prices
results in nearly identical forward three--month LIBORs, even
though the resulting instantaneous forward rates are very
different. This also implies that if one is only concerned with
rates such as forward three--month LIBORs, using loglinear
interpolation of zero coupon bond prices is for all intents and
purposes arbitrage free.

Introducing short bond volatility as per method 2 makes a
difference for the interpolated LIBORs only when the interest
rate term structure changes slope. The vertical gridlines in
figure \ref{LIBORTS} represent boundaries between the accrual
periods in the original discrete tenor. Interpolated forward
LIBORs with start date in an accrual period immediately preceding
a change in the slope of the term structure depart from the
linearly interpolated plots and curve ``outward''. This is due to
the fact that a forward LIBOR for a given start date $T$ is
interpolated using the forward LIBOR for the preceding and the
two following discrete tenor start dates $T_{\eta(T)-1}$,
$T_{\eta(T)}$ and $T_{\eta(T)+1}$, as can easily be seen by
appropriately inserting (\ref{corrfac}) into (\ref{ipLIBOR}).

\begin{figure}[tb]
\includegraphics[width=.99\textwidth]{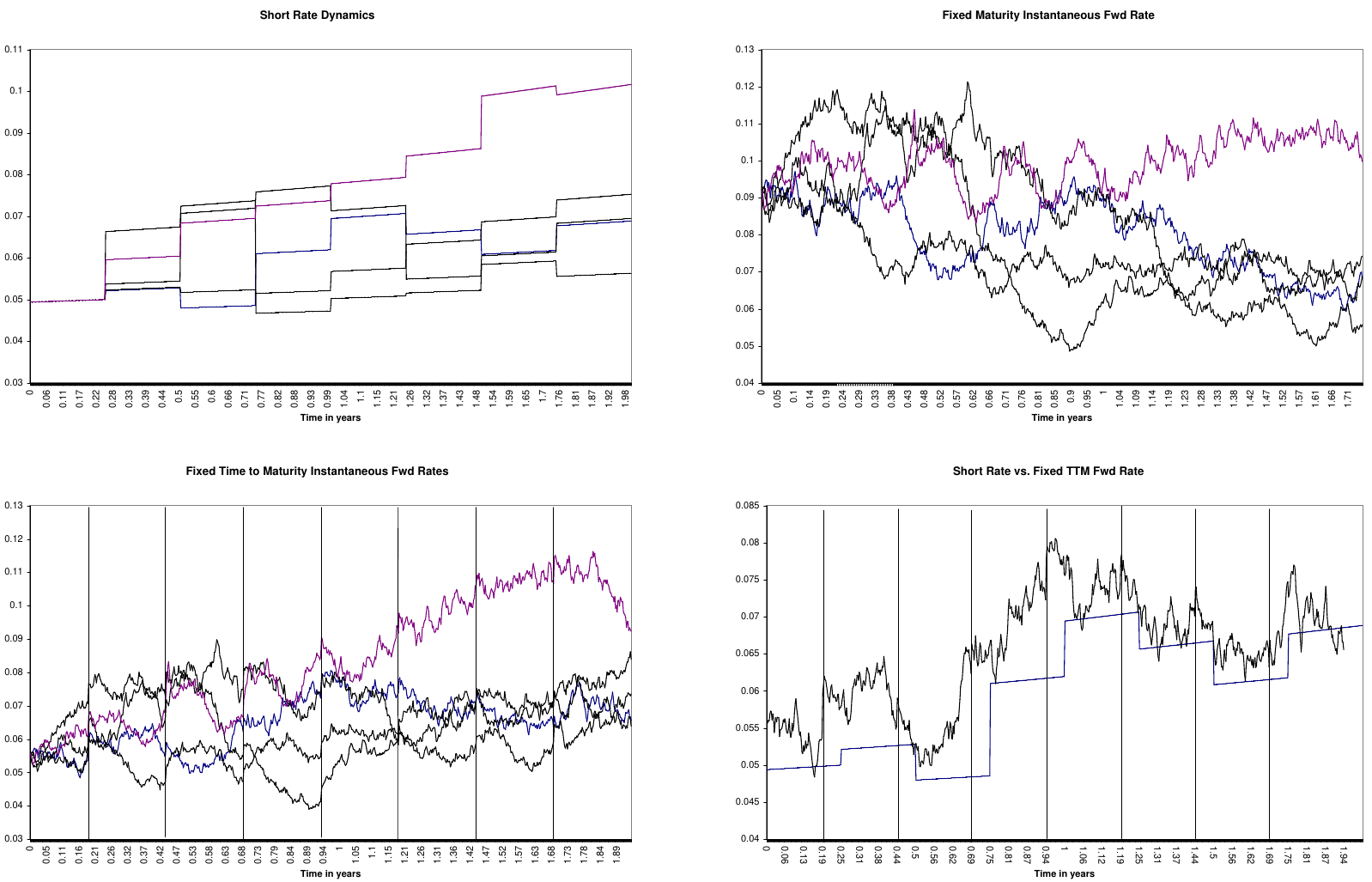}
\caption{Rate dynamics under interpolation by daycount fractions}\label{meth1dyn}
\end{figure}
\subsection{Rate Dynamics}
\begin{figure}[tb]
\includegraphics[width=.99\textwidth]{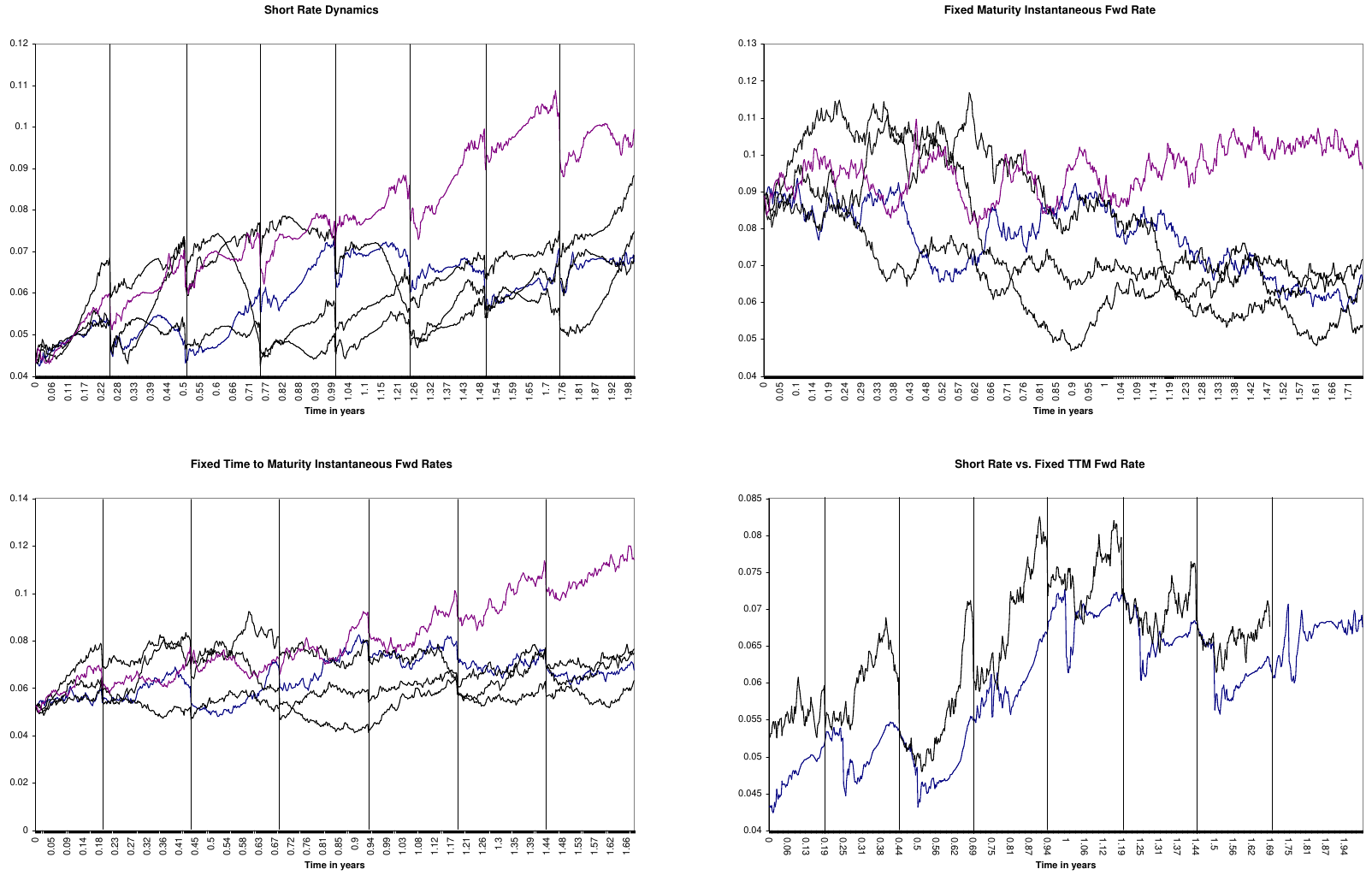}
\caption{Rate dynamics under interpolation with short bond
volatility}\label{meth2dyn}
\end{figure}
The dynamics which the two methods imply for interpolated rates
are illustrated in Figures \ref{meth1dyn} and \ref{meth2dyn}. Not
only, as in the previous section, do equations (\ref{meth1}) and
(\ref{meth2}) imply discontinuities in the instantaneous rates in
the maturity dimension, but also in the time dimension. This
becomes particularly clear in the dynamics of the continuously
compounded short rate under interpolation by daycount fractions:
This rate evolves deterministically within each accrual period of
the discrete tenor structure and jumps at the accrual period
boundaries. For instantaneous forward rates $f(t,T)$ the dynamics
are stochastic as long as the time to maturity $T-t$ is greater
than the accrual period length $\delta$. The maturity $T$ of the
forward rate determines which forward LIBORs are used in the
interpolation, so if one considers an instantaneous forward rate
for a fixed maturity, this rate will be interpolated from the same
forward LIBORs at all times. Since the LIBOR dynamics are
modelled to be continuous, the interpolated rate will not jump
(cf. plot in upper right of Figure \ref{meth1dyn}). On the other
hand, holding time to maturity (TTM) fixed means that there will
be jumps when the maturity $T$ moves across a boundary between
accrual periods in the original discrete tenor structure. The
bottom two plots in Figure \ref{meth1dyn}) illustrate this. The
time to maturity of the instantaneous forward rate shown here is
$1.25\delta$, resulting in a jump at three--quarters of every
$\delta$--interval of the tenor structure; these points are
marked by vertical gridlines in the plot. Thus different points
on the forward rate curve jump at different times.

The same observations can be made for the dynamics of
instantaneous rates under interpolation with short bond
volatility, as shown in Figure \ref{meth2dyn}. The only
qualitative difference is that now rates with times to maturity
less than $\delta$, including the continuously compounded short
rate, also have stochastic dynamics (cf. the upper left plot of
Figure \ref{meth2dyn}).

As in the previous section, the observed discontinuities apply
only to instantaneous rates. Of more practical relevance is the
fact that interpolated rates for accrual periods of any arbitrary
length, for example a $\delta$--compounded forward LIBOR $L(t,T)$
for a ``broken date'' maturity $T\not=T_{\eta(T)}$, will have
continuous dynamics: The instantaneous rates are continuous almost
everywhere under the Lebesgue measure on the time line, thus any
jumps are integrated out once one considers accrual periods of
greater than infinitesimal length.

\begin{figure}[tb]
\begin{minipage}[t]{0.445\textwidth}
\includegraphics[width=.99\textwidth]{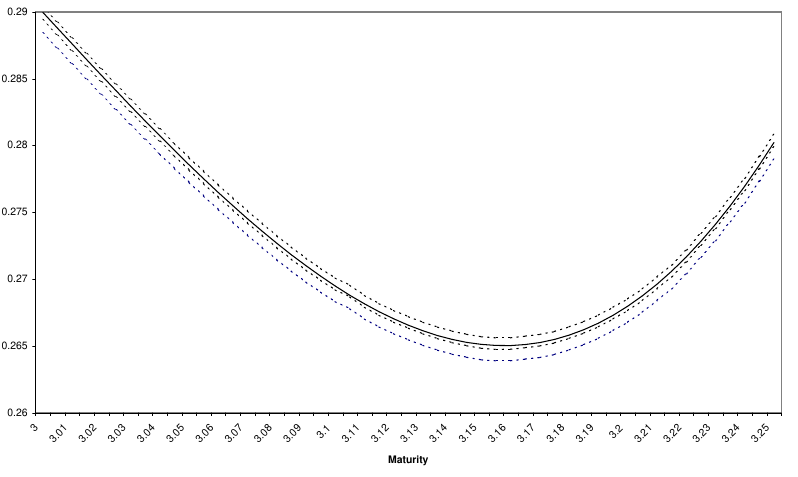} \caption{}\label{impvol1}
\end{minipage}\hfill\begin{minipage}[t]{0.445\textwidth}
\includegraphics[width=.99\textwidth]{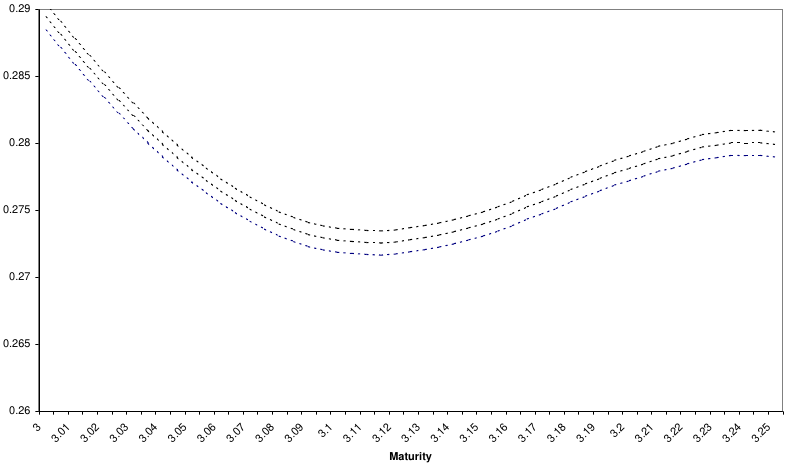} \caption{}\label{impvol2}
\end{minipage}
\begin{center}
Implied volatilities resulting from\\
interpolation by daycount
fractions vs. interpolation with short bond volatility
\end{center}
\end{figure}
\section{Broken Date Caplets: A Comparison}\label{cap}
Another point of interest is how the choice of interpolation
method affects the pricing of derivatives. In this context, the
volatility of interpolated rates is of key importance. When
extending the discrete tenor model to continuous tenor, one is
faced with a tradeoff of completing the model via volatilities or
interest rate interpolation. By fixing the volatilities of rates
for all maturities, say all $\delta$--compounded forward LIBORs
$L(\cdot,T)$, $T\in[0,T^*-\delta]$ as in MSS and BGM, the model is
completely specified and there is no need to interpolate interest
rates. On the other hand, when the interpolated rates are given as
functions of the discrete tenor forward LIBORs, their volatilities
can be calculated by an application of Ito's Lemma. The former
approach has the disadvantage of rather intractable dynamics for
rates other the $\delta$--compounded forward LIBORs, while latter
implies that the ``broken date'' forward LIBORs $L(t,T)$,
$T\not=T_{\eta(T)}$ are not lognormal. Applying Ito's Lemma to
(\ref{ipLIBOR}), the relative volatility of $L(t,T)$ is
\begin{multline*}
\vol[L(t,T)] = L(t,T)^{-1}\\
\cdot\sum_{i=0}^{N-1}\partial_{L(t,T_i)}\left(
\frac1{\delta}\left(\frac{B(t,T)}{B(t,T_{\eta(T)})} (1+\delta
L(t,T_{\eta(T)}))\frac{B(t,T_{\eta(T)+1})}{B(t,T+\delta)}-1\right)\right)
L(t,T_i)\lambda(t,T_i)\\
\end{multline*}
For method 1 this becomes
\begin{multline*}
\vol[L(t,T)] =
\frac1{\delta}L(t,T)^{-1}\left((T_{\eta(T)}-T)\frac{1+\delta
L(t,T_{\eta(T)})}{1+(T_{\eta(T)}-T)L(t,T_{\eta(T)})}
\lambda(t,T_{\eta(T)-1})L(t,T_{\eta(T)-1})\right.\\[2mm]
+(1+(T_{\eta(T)}-T)L(t,T_{\eta(T)-1}))\\[2mm]
\cdot\left.\frac{\delta(1+(T_{\eta(T)}-T)L(t,T_{\eta(T)}))
-(T_{\eta(T)}-T)(1+\delta
L(t,T_{\eta(T)}))}{(1+(T_{\eta(T)}-T)L(t,T_{\eta(T)}))^2}
\lambda(t,T_{\eta(T)})L(t,T_{\eta(T)})\right)\\
\end{multline*}
Figures \ref{impvol1} and \ref{impvol2} illustrate how the two
interpolation methods presented in the previous sections affect
the prices for ``broken date'' caplets. In each plot, we vary the
maturity $T$ of the forward LIBOR $L(t,T)$ underlying the caplet
within one $\delta$--accrual period of the discrete tenor
structure (i.e. the caplets on the two endpoints of each plot are
based on non-interpolated rates). The strike is chosen at 1.25
times the at--the--money level and the volatility function is
two--dimensional exponentially decaying.\footnote{For details see
appendix.} This choice should provide a representative example
well away from the trivial at--the--money, one--dimensional
constant volatility case. The caplet prices were generated by
Monte Carlo simulation; the middle line in each plot give the
Monte Carlo estimate, while the outside lines are the confidence
interval boundaries two standard deviations to either side of the
Monte Carlo estimate. One million MC runs make these confidence
bounds reasonably tight. We represent the caplet prices in terms
of their Black/Scholes implied volatility, i.e. in terms of the
one--dimensional constant relative volatility of the underlying
rate which would result in the same caplet price. The dip in the
implied volatilities is due to the fact that the interpolation of
$L(T,T)$ at maturity of the caplet is in part based on
$L(T_{\eta(T)-1},T_{\eta(T)-1})$, i.e. a rate which evolves
stochastically only until $T_{\eta(T)-1}<T$. This effect is less
pronounced under interpolation with short bond volatility, as in
this case $L(T_{\eta(T)-1},T_{\eta(T)-1})$ plays a lesser role.

The disadvantage that the ``broken date'' forward LIBORs are not
lognormal is greatly alleviated by the fact that their
distribution is actually very close to lognormal. One can apply
the argument which was first used to derive an approximate
swaption formula in a lognormal forward LIBOR model.\footnote{This
argument first appears in \mycite{OZ:BGM:97}. It was developed
further in \mycite{OZ:BDB:98} and formalized by
\mycite{OZ:BW:00}. This approximation works well not only for
pricing, but also for hedging, as demonstrated in
\mycite{OZ:DSB:00}.} Noting that the level dependence of the
volatility of the interpolated forward LIBORs varies slowly
compared to the rates themselves, one can derive an approximate
closed--form solution for the ``broken date'' caplet by
calculating the level dependence with respect to the initial
rates, e.g. for method 1 by setting
\begin{eqnarray*}
\volapprox[L(t,T)] &=&L(0,T)^{-1}\\
&& \cdot\left(\frac{\partial
L(0,T)}{\partial{L(0,T_{\eta(T)-1})}}L(0,T_{\eta(T)-1})\lambda(t,T_{\eta(T)-1})
+ \frac{\partial
L(0,T)}{\partial{L(0,T_{\eta(T)})}}L(0,T_{\eta(T)})\lambda(t,T_{\eta(T)})\right)\\
&=& \frac1{\delta}L(0,T)^{-1}\Biggl((T_{\eta(T)}-T)\frac{1+\delta
L(0,T_{\eta(T)})}{1+(T_{\eta(T)}-T)L(0,T_{\eta(T)})}
\lambda(t,T_{\eta(T)-1})L(0,T_{\eta(T)-1})\\[2mm]
&& +\ (1+(T_{\eta(T)}-T)L(0,T_{\eta(T)-1}))\\[2mm]
&& \qquad\cdot\frac{\delta(1+(T_{\eta(T)}-T)L(0,T_{\eta(T)}))
-(T_{\eta(T)}-T)(1+\delta
L(0,T_{\eta(T)}))}{(1+(T_{\eta(T)}-T)L(0,T_{\eta(T)}))^2}\\[2mm]
&&\qquad\qquad\cdot\lambda(t,T_{\eta(T)})L(0,T_{\eta(T)})\Biggr)
\end{eqnarray*}
The Black/Scholes implied volatility is then given by
$$
\frac1{\sqrt{T}}\sqrt{\int_0^T(\volapprox[L(s,T)])^2ds}
$$
\begin{eqnarray*}
&=& \frac1{\sqrt{T}}\Biggl(2\frac{\partial
L(0,T)}{\partial{L(0,T_{\eta(T)-1})}}L(0,T_{\eta(T)-1})
\frac{\partial
L(0,T)}{\partial{L(0,T_{\eta(T)})}}L(0,T_{\eta(T)})\cov(T_{\eta(T)-1},T_{\eta(T)})\\
&& +\ \left(\frac{\partial
L(0,T)}{\partial{L(0,T_{\eta(T)-1})}}L(0,T_{\eta(T)-1})
\overline{\lambda}(0,T_{\eta(T)-1})\right)^2\\
&& +\ \left(\frac{\partial
L(0,T)}{\partial{L(0,T_{\eta(T)})}}L(0,T_{\eta(T)})
\overline{\lambda}(0,T_{\eta(T)})\right)^2\Biggr)^{\frac12}
\end{eqnarray*}
with
\begin{eqnarray*}
\cov(T_{\eta(T)-1},T_{\eta(T)}) &=&
\int_0^{T_{\eta(T)-1}}\lambda(s,T_{\eta(T)-1})\lambda(s,T_{\eta(T)})ds\\
\overline{\lambda}(0,T_{\eta(T)-1}) &=&
\sqrt{\int_0^{T_{\eta(T)-1}}\lambda^2(s,T_{\eta(T)-1})ds}\\
\overline{\lambda}(0,T_{\eta(T)}) &=&
\sqrt{\int_0^{T_{\eta(T)}}\lambda^2(s,T_{\eta(T)})ds}
\end{eqnarray*}
As illustrated by Figure \ref{impvol1}, the resulting
approximation of Black/Scholes implied volatilities is very
accurate.

\section{Conclusion}\label{end}
Extending the models of market observable interest rates from
discrete to continuous tenor by interpolation is particularly
useful in implementations where there are some financial products
which must be priced numerically, as the Markovian structure of
the discrete tenor model is preserved. It is important to note
that the interpolation method cannot be chosen arbitrarily for
all maturities. Rather, it must take into account the relevant
no--arbitrage conditions.

Shifting the focus from instantaneous forward rates to market
observables such as forward LIBOR, the rate dynamics implied by
the proposed interpolation methods are reasonable. These methods
supply alternatives to the continuous tenor versions of the
lognormal forward rate Market Models proposed in the literature
and in those cases where moving to continuous tenor by
interpolation entails a loss of tractability, for example for
``broken date'' caplets, very accurate approximations exist.

\section*{Appendix: Model Parameters}
To generate the plots, two volatility specifications $\lambda_1$
and $\lambda_2$ were used:
\begin{eqnarray*}
\lambda_1(t,T) &=& 0.3 \qquad\qquad\qquad\qquad\quad\forall\ t<T<T^*\\
\lambda_2(t,T) &=& \left(\begin{array}{c} 0.6e^{-0.8(T-t)}\\
0.1e^{-0.01(T-t)} \end{array}\right) \qquad\ \forall\ t<T<T^*
\end{eqnarray*}
The length $\delta$ of the discrete tenor accrual period is in all cases 0.25.

\begin{table}[tb]
\caption{}\label{params} \centerline{\begin{tabular}{|c|c|c|c|c|c|}
\hline
       & Initial   &            &         &          & Fixed \\
Figure & Term      & Volatility & $T^*$    & Fixed    & Time to \\
       & Structure &            &         & Maturity & Maturity \\
\hline
\ref{daycountTS} & 1 & $\lambda_1$ & 10   & NA       & NA \\
\ref{shortvolTS} & 1 & $\lambda_1$ & 10   & NA       & NA \\
\ref{LIBORTS}    & 2 & $\lambda_1$ & 10   & NA       & NA \\
\ref{meth1dyn}   & 3 & $\lambda_1$ & 2.25 & 1.8125   & 0.3125 \\
\ref{meth2dyn}   & 3 & $\lambda_1$ & 2.25 & 1.8125   & 0.3125 \\
\ref{impvol1}    & 3 & $\lambda_2$ & 4.25 & NA       & NA \\
\ref{impvol2}    & 3 & $\lambda_2$ & 4.25 & NA       & NA \\
\hline
\end{tabular}}
\end{table}
Three initial term structures of discrete tenor forward LIBORs
are considered:
\begin{enumerate}
\item Linearly increasing from 4\% at 0 years up to 6\% at 5 years and then
linearly decreasing down to 4\% at 10 years.
\item Linearly increasing from 5\% at 0 years up to 6.7\% at 4.25 years, then
linearly decreasing to 6.5\% at 4.75 years, then linearly increasing to 6.8\% at
5.5 years, and then linearly decreasing down to 5\% at 10 years.
\item Linearly increasing from 5\% at 0 years up to 10\% at $T^*-\delta$.
\end{enumerate}

\bibliography{amf,oz,statneu,zusatz,jfin,pubinst,FMMA}
\end{document}